\documentclass{article}
\usepackage[preprint,nonatbib]{neurips_2019}
\pdfoutput=1

\usepackage{alltt}
\usepackage{color}
\usepackage{colortbl}
\usepackage[utf8]{inputenc} 
\usepackage[T1]{fontenc}    
\usepackage{hyperref}       
\usepackage{url}            
\usepackage{booktabs}       
\usepackage{amsfonts}       
\usepackage{nicefrac}       
\usepackage{microtype}      
\usepackage{tikz}
\usepackage{wrapfig}
\usepackage{xspace}
\usepackage{todonotes}

\newcommand*{\lale}{\textsc{Lale}\xspace}
\newcommand*{\ckm}{\checkmark}
\newcommand*{\python}[1]{\texttt{\small #1}}
\newcommand*{\pipe}{\python{>{}>}\xspace}
\newcommand*{\choice}{\python{|}\xspace}
\newcommand*{\union}{\python{\&}\xspace}

\definecolor{metamodel}{HTML}{6CA6CD} 
\definecolor{planned}{HTML}{7EC0EE} 
\definecolor{trainable}{HTML}{B0E2FF} 

\title{Type-Driven Automated Learning with \lale}

\author{%
  Martin Hirzel, Kiran Kate, Avraham Shinnar, Subhrajit Roy, Parikshit Ram\\
  IBM Research\\
  \textsf{\small\{hirzel,kakate,shinnar\}@us.ibm.com,subhrajit.roy@au1.ibm.com,parikshit.ram@ibm.com}}

\begin{document}

\maketitle

\begin{abstract}
  Machine-learning automation tools, ranging from humble grid-search to
hyperopt, auto-sklearn, and TPOT, help explore large search spaces of
possible pipelines. Unfortunately, each of these tools has a
different syntax for specifying its search space, leading to lack of
portability, missed relevant points, and spurious points that
are inconsistent with error checks and documentation of the searchable
base components. This paper proposes using types (such as enum, float,
or dictionary) both for checking the correctness of, and for
automatically searching over, hyperparameters and pipeline configurations.
Using types for both of these purposes
guarantees consistency. We present \lale, an embedded language
that resembles scikit learn but provides better automation,
correctness checks, and portability. \lale extends the reach of
existing automation tools across
data modalities (tables, text, images, time-series)
and programming languages (Python, Java, R). Thus, data scientists can
leverage automation while remaining in control of
their work.

\end{abstract}

\section{Introduction}\label{sec:introduction}

When machine-learning practitioners assemble and configure pipelines of data
transformations and machine-learning models, they face an
over-abundance of choices. Fortunately, there are tools such as
GridSearchCV~\cite{buitinck_et_al_2013},
hyperopt~\cite{bergstra_et_al_2015},
auto-sklearn~\cite{feurer_et_al_2015}, and
TPOT~\cite{olson_et_al_2016} that automatically search over a space of
such choices for combined algorithm selection and hyperparameter
optimization (CASH). This paper presents a novel type-driven approach for
making CASH tools easier to use correctly and more portable, thus
expanding the set of people who can benefit from data science
automation.

Unfortunately, previous CASH tools each have their
own syntax for specifying the search space, which the user must
learn. In practice, the search space often includes invalid points,
causing the underlying library to report an error or to
produce useless results. To prevent that, users often overcorrect by
considering only an obviously valid subset of the search space, at
the risk of missing relevant points. As the search space specification
is a code artifact that is separate from the documentation or
error-checks of the underlying library, maintaining it is labor
intensive and brittle. This is exacerbated by the fact that to use
multiple CASH tools, one must specify the same
search space multiple times. One work-around is to only use
specifications pre-bundled with the tool, but that limits the available
operators (transformers or estimators).

The goals of this paper are automation together with usability and
portability. For automation, we aim to augment but not replace the
data scientist, by letting them bind some free variables of their
problem space by automated search and others by hand.
We refer to this as \emph{lifecycle as bindings}, because binding free
variables (e.g., configuring hyperparameters) transitions a pipeline
to the next lifecycle state (e.g., makes it trainable).
For usability, we aim to retain the familiar interfaces of
popular libraries while improving correctness by embracing the DRY
(don't repeat yourself) principle~\cite{hunt_thomas_1999}.
And for portability, we offer a pluggable middleware that is
independent of any particular machine-learning library or CASH tool
and even works seamlessly across different programming languages.

This paper introduces \lale, an implementation of our type-driven
learning automation approach. \lale (Language for Automated Learning
Exploration) is a new domain-specific embedded language
(DSEL,~\cite{hudak_1998}) designed around JSON
Schema~\cite{pezoa_et_al_2016} and scikit
learn~\cite{buitinck_et_al_2013}. As a DSEL, \lale tries to combine
the best qualities of a library and a language. On the one hand, \lale
is a pip-installable Python library so users can edit code in familiar
Python syntax using Python tooling such as Jupyter, MyPy, or IDEs.
On the other hand, \lale includes a
compiler that auto-generates search spaces for CASH tools; \lale
provides pipeline combinators that guarantee consistency across the
lifecycle; and \lale avoids mutable state and magic strings, making it
more robust.

\lale is compatible with scikit learn~\cite{buitinck_et_al_2013} in an
attempt to capitalize on its many good qualities, including
familiarity, ease of use, clear basic concepts, many operators, wide
adoption, and interoperability with other libraries.  \lale search
spaces and \lale pipelines can include operators written in Python,
Java, and R. \lale uses JSON
Schema~\cite{pezoa_et_al_2016}, a type system for JSON, for type
annotations on hyperparameters.  We found JSON Schema to be
well-suited to capture the intricacies of hyperparameters including
categorical and continuous values and conditional
dependencies. Furthermore, JSON Schema is widely adopted and on a path
to standardization. There are JSON Schema validators in Python,
JavaScript, Java, and several other programming languages.
Furthermore, JSON Schema is closely linked to Swagger for specifying
web APIs, which can be used to serve data science pipelines.

The contributions of this paper are:

\begin{itemize}
  \item Lifecycle as bindings. \lale lets users bind some free
    variables of their problem space by automated search and others by
    hand. These bindings in an operator determine which state of its
    lifecycle it is in, thus giving users fine-grained control and
    transparency (Section~\ref{sec:lifecycle}).
  \item Types as search spaces. \lale uses types, encoded using JSON
    Schema, as a single source of truth not just for correctness
    checks (the traditional purpose of types) but also for automation
    (by novel compilers to tool-specific search spaces)
    (Section~\ref{sec:searchspaces}).
  \item Scikit learn-compatible portability. \lale adopts a novel
    technique for simultaneously being compatible with scikit learn
    but also more portable, allowing seamless interoperability between
    machine learning and deep learning and between Python, Java, and R
    (Section~\ref{sec:portability}).
\end{itemize}

This paper demonstrates \lale with case studies from four different
data modalities: tables, text, images, and time-series.  This
generality of modality demonstrates \lale's portability.  Overall, we
argue that our type-driven approach makes machine learning more automated,
usable, and portable.

\section{Related Work}\label{sec:related}

Combined algorithm selection and hyperparameter optimization (CASH)
requires a library of data science operators (transformers or
estimators). Popular such libraries include scikit
learn~\cite{buitinck_et_al_2013}, Weka~\cite{hall_et_al_2009},
R~\cite{ihaka_gentleman_1996}, pandas~\cite{mckinney_2011},
Keras~\cite{chollet_2015}, and Spark MLlib~\cite{meng_et_al_2016}.
Most of them have high-quality human-readable documentation but lack
machine-readable specifications of the induced search spaces.  Thus,
such specifications are left to individual CASH tools. Weka stands out
by specifying hyperparameter types but does not specify conditional
hyperparameter dependencies. \lale currently wraps operators from
scikit learn, Weka, and R, and provides detailed machine-readable
specifications for them.

\begin{table}[b]
\caption{\label{tab:expressiveness}Search space specification expressiveness of CASH tools.}
\centerline{\begin{tabular}{@{}lcccccccl@{}}
                      &  cat & cont & dict & $\vee$ & $\wedge$ & $\neg$ & cond & nesting\\\cmidrule{2-9}
  GridSearchCV        & \ckm &      & \ckm & \ckm   &          &        &      & $\vee(\textrm{dict}\{\textrm{cat}^*\}^*)$\\
  RandomizedSearchCV  & \ckm & \ckm & \ckm &        &          &        &      & $\textrm{dict}\{\textrm{cat}^*,\textrm{cont}^*\}$\\
  auto-sklearn (SMAC) & \ckm & \ckm & \ckm &        &          &        & \ckm & $\textrm{dict}\{\textrm{cat}^*,\textrm{cont}^*\}\wedge\textrm{cond}^*$\\
  hyperopt            & \ckm & \ckm & \ckm & \ckm   &          &        &      & fully nested\\
  TPOT                & \ckm &      & \ckm & \ckm   &          &        &      & $\vee(\textrm{dict}\{\textrm{cat}^*\}^*)$\\
  \lale (JSON Schema) & \ckm & \ckm & \ckm & \ckm   & \ckm     & \ckm   &      & fully nested
\end{tabular}}
\end{table}

Most CASH tools expose a syntax for specifying their search space, and
Table~\ref{tab:expressiveness} compares how expressive that is.
Each point in a search space is a pipeline configured with its
operator choices and hyperparameter values. For
ease of comparison, we look past superficial choices of identifiers
and symbols to the underlying concepts: categoricals (cat),
continuous (cont), dictionary (dict), Boolean connectives
($\vee$, $\wedge$, $\neg$), and conditional (cond). The final
column summarizes how these can be nested.

\emph{Grid search} is a humble but effective way to explore a search
space.  Many data science libraries implement it; here, we
focus on GridSearchCV from scikit learn~\cite{buitinck_et_al_2013}.
It supports search spaces of the form
\mbox{$\vee(\textrm{dict}\{\textrm{cat}^*\}^*)$}: a top-level
disjunction of dictionaries of categorical hyperparameters. Continuous
hyperparameters must be discretized for use with GridSearchCV. Grid
search has the advantage of not getting stuck in local minima, but
may miss relevant points due to discretization, and it is
slow for large grids. Most people use
it only for hyperparameter tuning (not algorithm selection). 
The \lale compiler has a backend
that turns GridSearchCV into a full-fledged CASH tool.

\emph{Randomized search}, like grid search, cannot get stuck in local
minima. In scikit learn, RandomizedSearchCV, unlike GridSearchCV,
lacks Boolean connectives such as a $\vee$ necessary for handling
algorithm selection or conditional hyperparameters, making it
insufficient for full-fledged CASH.

\emph{SMAC} (Sequential Model-based Algorithm Configuration) is one
way to approximate the probabilistic dependency of the loss on the
search space point, using an acquisition function to balance
exploration against exploitation~\cite{hutter_hoos_leytonbrown_2011}.
SMAC is used internally by both auto-Weka~\cite{thornton_et_al_2013}
and auto-sklearn~\cite{feurer_et_al_2015}. Its search spaces have the
form \mbox{$\textrm{dict}\{\textrm{cat}^*,\textrm{cont}^*\}\wedge\textrm{cond}^*$},
so it supports conditional
hyperparameters directly and algorithm selection indirectly via
synthetic indicator hyperparameters.  The \lale compiler has a SMAC backend that
takes advantage of these.

\emph{Hyperopt}~\cite{bergstra_et_al_2015}, and consequently
hyperopt-sklearn~\cite{komer_bergstra_eliasmith_2014}, uses TPE
(Tree-structured Parzen Estimator~\cite{bergstra_et_al_2011}) to
model the loss across the search space. Its search
spaces support $\vee$, thus enabling algorithm selection and conditional
hyperparameters. The \lale compiler also has a backend for hyperopt.

\emph{TPOT} (Tree-Based Pipeline Optimization
Tool~\cite{olson_et_al_2016}) uses genetic algorithms. Continuous
hyperparameters must be discretized for use with TPOT. In essence, its
search space specification has the same form as in GridSearchCV, even
though at the surface, its syntax differs. Unlike the other CASH tools
discussed here, TPOT can explore new graph topologies for pipelines.

In contrast to scikit learn's GridSearchCV or RandomizedSearchCV,
SMAC, hyperopt, or TPOT, \lale uses JSON
Schema~\cite{pezoa_et_al_2016} to specify search spaces. JSON Schema,
normally used for correctness checks, turns out to be an 
expressive basis for specifying constrained search spaces.  Furthermore, it is
independent of individual machine learning libraries, CASH tools, or
programming languages.

One focus of \lale is \emph{portable} learning automation: \lale
enables off-the-shelf CASH tools to discover pipelines that mix
operators from Python, Java, and~R. In earlier work, PMML provided an
interchange format to export and import machine learning operators across
libraries and languages~\cite{guazzelli_et_al_2009}. Spark MLlib is
available via APIs in different languages~\cite{meng_et_al_2016}.  And
Kubeflow Pipelines are designed for portable machine learning
workflows based on Docker containers~\cite{kubeflow_authors_2018}.  In
contrast to PMML, Spark MLlib, or Kubeflow Pipelines, \lale is
designed to facilitate automation. In future work, \lale
may offer interoperability with these technologies to improve scaling.


\section{Lifecycle as Bindings}\label{sec:lifecycle}

Machine learning pipelines have their own \emph{lifecycle}.  For CASH, the
relevant lifecycle transitions encompass selecting operators from a
library and arranging them into candidate pipelines; tuning
hyperparameters; and finally training and evaluating candidates to
find the best. We wanted to give data scientists fine-grained control
over which transitions to do manually and which to automate. However,
this posed a DSEL design challenge: how can \lale offer a unified
experience for the manual and automated tasks? Such as a unified
experience is crucial to ensure a pipeline remains consistent
throughout its lifecycle and to help the data scientist understand the
results of automation.

\lale offers three \emph{combinators} for arranging operators into
pipelines: \pipe, \union, and \choice. The \pipe combinator behaves
like scikit learn's \python{make\_pipeline}. The \union combinator
behaves like scikit learn's \python{make\_union}, except without
concatenating the features at the end. Instead, \lale provides a
separate \python{ConcatFeatures} operator to increase flexibility.
The \choice combinator
implements operator choice and does not have an equivalent in scikit
learn.  Users can also write \choice as \python{make\_choice}.

\begin{wrapfigure}{r}{0pt}
\vspace*{-5mm}
\includegraphics[scale=0.4]{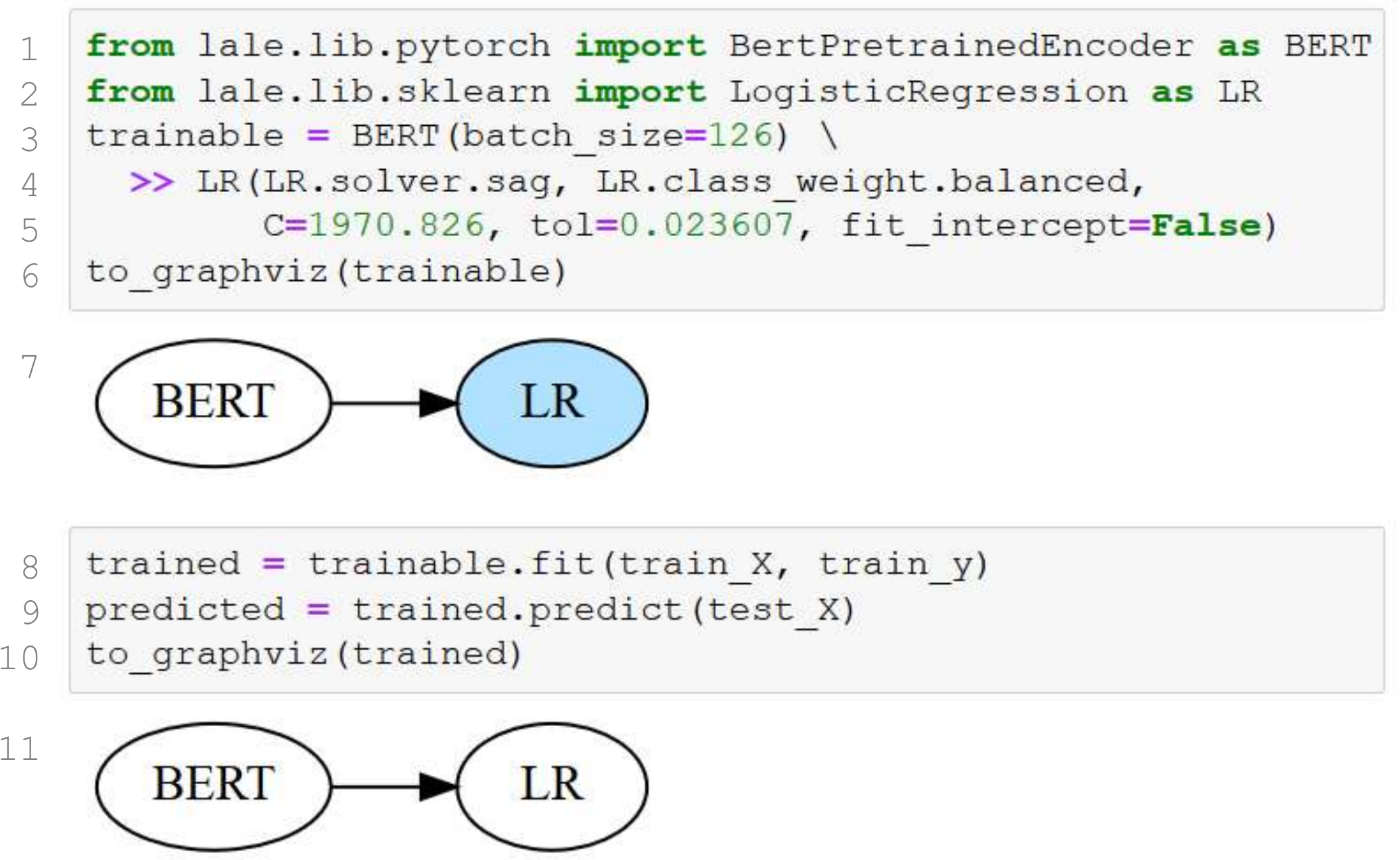}
\caption{\label{fig:lifecycle_notebook_manual}Manual text processing pipeline example.}
\vspace*{-5mm}
\end{wrapfigure}

Figure~\ref{fig:lifecycle_notebook_manual} illustrates how to perform
lifecycle transitions manually for a text classification example.
Line~1 imports BERT~\cite{devlin_et_al_2018}, a text embedding based
on neural networks, in this case pre-trained and implemented in
PyTorch~\cite{paszke_et_al_2017}. Line~2 imports the scikit-learn
logistic regression as \python{LR}. \mbox{Lines 3--5}
configure the hyperparameters of both \python{BERT} and \python{LR} and arrange them
into a pipeline using the \pipe combinator. Note the argument
\python{LR.solver.sag} is a Python enumeration, which \lale
auto-generates from the same JSON schema used for search spaces.
\mbox{Lines 6--7} visualize the computational graph. \lale reflects
the name used in the source code in the visualization, e.g., \python{LR} instead of
\python{LogisticRegression}. Node colors indicate lifecycle states,
here white for \emph{trained} and light blue for \emph{trainable}. It
is a static error to call \python{predict} or \python{transform} on an
operator unless it is trained. Line~8 trains the pipeline, resulting
in a new pipeline \python{trained} that is distinct from the original
\python{trainable}.  Keeping these in separate Python variables and
objects makes it easier to track lifecycle states and should also
help parallel or cloud-based execution and prevent
accidental overwrite, e.g., during $k$-fold cross validation.
Calling \python{predict} in Line~9 is valid. \mbox{Lines 10--11}
show that \python{trained} is fully trained (all white) and
consistent with \python{trainable} (same graph topology).

\begin{wrapfigure}{r}{0pt}
\begin{minipage}{.67\textwidth}
\vspace*{-4mm}
\includegraphics[width=\textwidth]{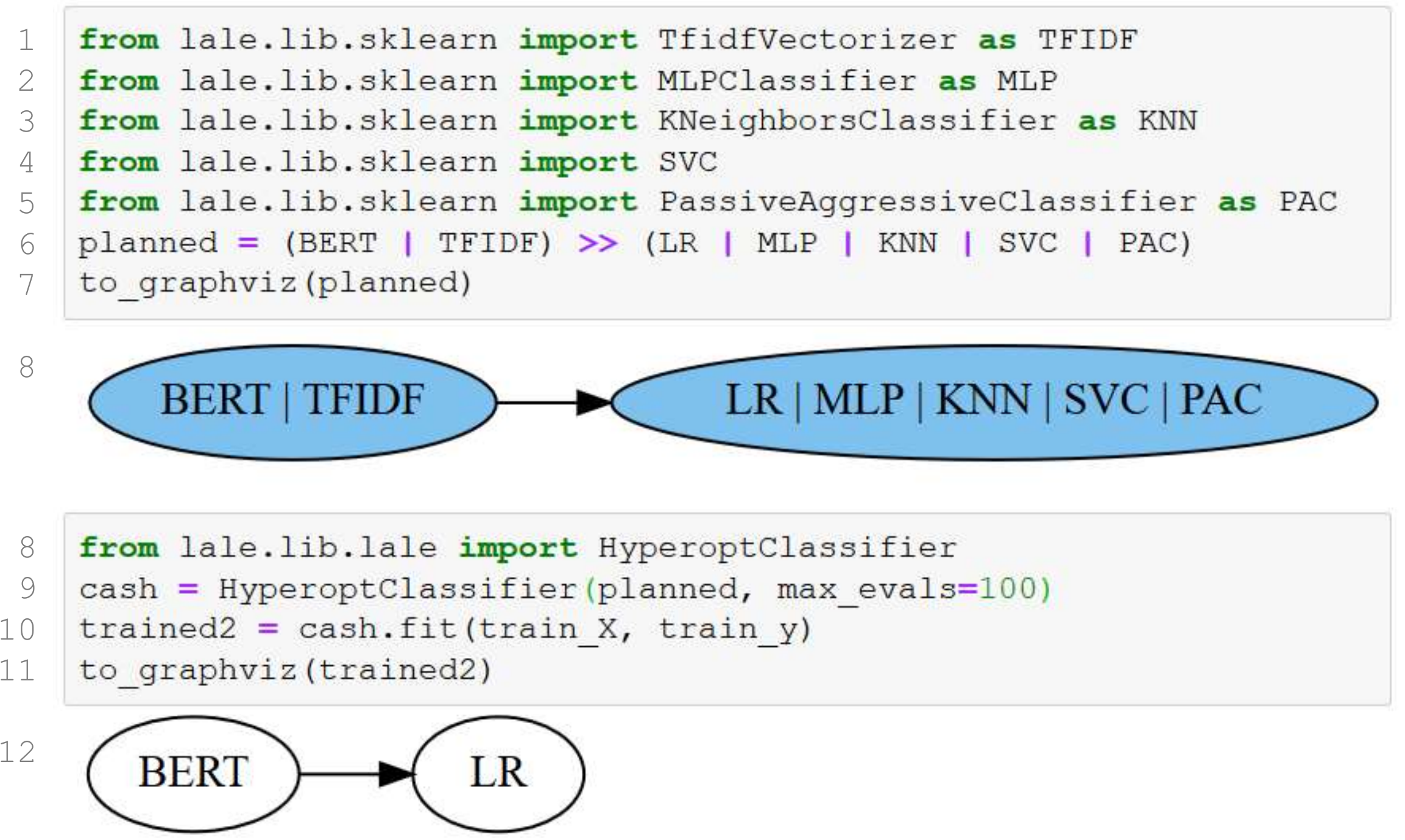}
\caption{\label{fig:lifecycle_notebook_auto}Automated text processing pipeline example.}
\vspace*{-4mm}
\end{minipage}
\end{wrapfigure}
Figure~\ref{fig:lifecycle_notebook_auto} automates
lifecycle transitions for this example.
\mbox{Lines 1--5} import a few more operators. Line~6 arranges them in
a pipeline. But unlike the earlier example, it does not manually bind
all operator selections, nor does it manually bind hyperparameters.
We refer to the lifecycle state where these properties are still free
(not yet
bound) as \emph{planned}. \mbox{Lines 7--8} visualize the computational
graph, rendering both steps in a darker blue to indicate their planned
state. Together with the hyperparameter schemas of individual
operators, a planned pipeline induces a search space. It is a static
error to call \python{fit} on an operator unless it is trainable.
Line~8 imports a CASH tool, and Line~9 instantiates it for our planned
pipeline. Line~10 runs the CASH tool to bind the free properties of
\python{planned}, resulting in \python{trained2}.
\mbox{Lines 11-12} demonstrate visually that the end
result is fully trained and has bound operator choices
consistent with the original planned pipeline.

\begin{wraptable}{r}{3.5in}
\vspace*{-6mm}
\caption{\label{tab:lifecycle_bindings}Properties bound by lifecycle state.}
\centerline{\begin{tabular}{@{}lll@{}}
             & Individual operator   & Pipeline\\\cmidrule{2-3}
  Meta-model & \cellcolor{metamodel}schemas, tags, priors & \cellcolor{metamodel}composable elements\\
  Planned    & \cellcolor{planned} & \cellcolor{planned}graph topology\\
  Trainable  & \cellcolor{trainable}hyperparameters & \cellcolor{trainable}operator choices\\
  Trained    & learned coefficients & \\
\end{tabular}}
\vspace*{-4mm}
\end{wraptable}
Table~\ref{tab:lifecycle_bindings} summarizes the lifecycle concepts.
States are ordered from top to bottom.  An operator is in a given
state if all properties up to that state are already bound. Properties
in later states may still be free. In other words, each lifecycle
state has a superset of the bound properties of its predecessor
state. Each method, such as \python{fit} or \python{predict}, requires
certain bound properties, such as hyperparameters or learned
coefficients. Therefore, each lifecycle state also supports a superset
of the methods of its predecessor.  Pretrained methods, such as
pretrained BERT, have no-op training. The state of an entire pipeline is the
least upper bound of the states of its steps. CASH tools may automate
one or multiple state transitions. For instance, TPOT can bind the
graph topology~\cite{olson_et_al_2016} whereas hyperopt uses a
pre-specified topology~\cite{bergstra_et_al_2015}.  \lale also
supports partial bindings, for instance, setting some hyperparameters
manually and others automatically, or transfer learning where some
learnable coefficients are pretrained and frozen.

Implementing \lale required some innovation in domain-specific
embedded languages (DSELs~\cite{hudak_1998}). First, we wanted to get
the benefits of static typestate checking~\cite{strom_yemini_1986},
such as error message or auto-complete proposals, but with
off-the-shelf Python tooling. We accomplished this by making the
lifecycle states manifest in the code as classes and making each state
a subclass of its predecessor.  For instance, \python{Trainable} is a
subclass of \python{Planned} and adds a \python{fit} method.  In
addition, we added Python~3 type annotations to guide error checking
tools. For instance, the return type of \python{fit} is
\python{Trained}.

One problematic case was omitting the \python{predict} or
\python{transform} methods from \python{Trainable}. Scikit learn
allows users to call first \python{fit} and then \python{predict} or
\python{transform} on the same object (of the same class). To ease
adoption, we took a softer approach: we gave \python{Trainable} a
\python{predict} or \python{transform} method but made it deprecated.
That way, users are encouraged but not forced to adopt a coding style
that cleanly separates lifecycle states. Users can even disable the
deprecation warning if they so desire.

The \lale combinators \pipe, \union, and \choice work uniformly at all
lifecycle states, from planned
(Figure~\ref{fig:lifecycle_notebook_auto}) to trainable and (pre-)trained
(Figure~\ref{fig:lifecycle_notebook_manual}). Getting this consistent
experience to work posed a challenge: whereas scikit learn implements the
trainable and trained states via instance methods, it implements the
planned state as a class, not an instance. For \lale, this was problematic,
because Python cannot overload \pipe, \union, and \choice as class
methods. Our solution was to consistently implement all lifecycle
states via instance methods. One special case is that scikit learn
sets hyperparameters with a constructor such as
\mbox{\python{BERT(batch\_size=126)}}; to turn this into an instance
method, \lale overloads \python{Planned.\_\_call\_\_}.

Finally, we wanted to make it easy for operator developers to
contribute operators to \lale. They should only have to write 
one class per operator, not one class per state per operator. Also,
they should only have to write scikit learn style constructors, not
overload \python{\_\_call\_\_}. \lale solves this by applying
state wrappers around contributed operator implementations. The
wrappers follow the hierarchy discussed before, creating new instances
of the wrapped implementation classes as needed.


\section{Types as Search Spaces}\label{sec:searchspaces}

This section describes how to translate from a \lale search space to
search spaces for three popular CASH tools: scikit learn's
GridSearchCV~\cite{buitinck_et_al_2013},
SMAC~\cite{hutter_hoos_leytonbrown_2011}, and
hyperopt~\cite{bergstra_et_al_2015}.  These are the most interesting
cases from Table~\ref{tab:expressiveness}; we omitted
RandomizedSearchCV because it is not expressive enough and
TPOT~\cite{olson_et_al_2016} because its search space specification
has the same form as that of GridSearchCV.  A \lale search space is
induced by a \lale pipeline (which may involve the choice
combinator~`\choice') alongside the JSON schemas for the hyperparameters
of each step in the pipeline. To avoid missing relevant points and to
avoid including invalid points, these schemas frequently include
constraints for conditional hyperparameters. Our compiler needs to
preserve the search space including constraints.

As a running example, consider the \lale pipeline
\mbox{\python{PCA >{}> (J48 | LR)}}, where \python{PCA} and
\python{LR} are the principal component analysis and logistic
regression from scikit learn and J48 is a decision tree with pruning
from Weka~\cite{hall_et_al_2009}.
These operators have many hyperparameters and constraints and \lale
handles all of them. For didactic purposes, this section discusses only
a representative subset:

\[\begin{array}{l@{\,:\;}l}
  \textit{PCA}
& \textrm{dict}\{N{:\,}(0..1) \vee [\textit{mle}])\}
\\
  \textit{J48}
& \textrm{dict}\{
    R{:\,}[\textit{true},\textit{false}],
    C{:\,}(0..1)\} \wedge
  ( \textrm{dict}\{R{:\,} \neg[\textit{true}]\} \vee
    \textrm{dict}\{C{:\,}[0.25]\})
\\
  \textit{LR}
& \textrm{dict}\{
    S{:\,}[\textit{linear}, \textit{sag}, \textit{lbfgs}],
    P{:\,}[\textit{l1}, \textit{l2}]\} \wedge
  ( \textrm{dict}\{S{:\,} \neg [\textit{sag}, \textit{lbfgs}]\} \vee
    \textrm{dict}\{P{:\,}[\textit{l2}]\})
\end{array}\]

The number of
components for PCA is given by $N$, which can be a continuous value
in~$(0..1)$ or the categorical value \textit{mle}. J48 has a
categorical hyperparameter $R$ to enable reduced error pruning and a
continuous confidence threshold $C$ for pruning; the side constraint
indicates that when $R$ is true then $C$ must \mbox{be 0.25}. LR has
two categorical hyperparameters $S$ (solver) and $P$ (penalty); the
side constraint indicates that solvers \textit{sag} and \textit{lbfgs}
only support penalty \textit{l2}.

\lale's search space compiler has two phases: normalizer and backend.
The normalizer transforms the schemas of individual operators
separately. The backend combines the schemas for the entire pipeline
and generates a search space in the format required by a given CASH
tool.

The \emph{normalizer} processes the schema for an individual operator
in a bottom-up pass. The desired end result is a search space in
\lale's \emph{normal form}, which is
\mbox{$\vee(\textrm{dict}\{\textrm{cat}^*,\textrm{cont}^*\}^*)$}.  At
each level, the normalizer simplifies children and hoists
disjunctions up.
\emph{Simplification} applies several semantics-preserving rewrites to
keep the size of the search space specification manageable and to
reduce the burden on the tool-specific backends, for instance:

\[\begin{array}{c}
   s_0 \vee \bot \;\Rightarrow\; s_0
    \qquad\qquad  s_0 \wedge \top \;\Rightarrow\; s_0
    \qquad\qquad \neg(s_0 \vee s_1) \;\Rightarrow\; (\neg s_0) \wedge (\neg s_1)\\
  {[}\textrm{cat}_0] \wedge {[}\textrm{cat}_1] \;\Rightarrow\; {[}\textrm{cat}_0
                                                 \cap \textrm{cat}_1]
                                                 \qquad\qquad
  {[}\textrm{cat}_0] \wedge \neg{[}\textrm{cat}_1] \;\Rightarrow\; {[}\textrm{cat}_0 \setminus \textrm{cat}_1]\\

    \textrm{dict}\{k_0: s_0, k_1: s_1\} 
    \wedge \textrm{dict}\{k_0: s_0', k_1: s_1\} \;\Rightarrow\; \textrm{dict}\{k_0: s_0 \wedge s_0', k_1: s_1\}\\
\end{array}\]

\emph{Disjunction hoisting} moves $\vee$ up to the top-level, using
semantics-preserving rewrites such as:

\[\begin{array}{r@{\;\Rightarrow\;}l}
  (s_0 \vee s_1) \wedge (s_2 \vee s_3) & (s_0 \wedge s_2) \vee (s_0 \wedge s_3) \vee (s_1 \wedge s_2) \vee (s_1 \wedge s_3)\\
  \textrm{dict}\{k_0: s_0 \vee s_0', k_1: s_1\} & \textrm{dict}\{k_0: s_0, k_1: s_1\} \vee \textrm{dict}\{k_0: s_0', k_1: s_1\}\\
\end{array}\]

The normalizer always terminates, making a single bottom-up
pass.  Its output specifies the same search space as
the input since the individual rewrites preserve semantics. The
normalizer is not guaranteed to reach normal form, but we test that it
does so for
all operators in the \lale
library. It also may generate redundant choices.  The resulting normalized schemas for our running example are:

\[\begin{array}{l@{\,:\;}l}
  \textit{PCA}
& \textrm{dict}\{N{:\,}(0..1)\} \vee
  \textrm{dict}\{N{:\,}[\textit{mle}]\}
\\
  \textit{J48}
& \textrm{dict}\{R{:\,}[\textit{false}],               C{:\,}(0..1)\} \vee
  \textrm{dict}\{R{:\,}[\textit{true},\textit{false}], C{:\,}[0.25]\}
\\
  \textit{LR}
& \textrm{dict}\{S{:\,}[\textit{linear}],                               P{:\,}[\textit{l1}, \textit{l2}]\} \vee
  \textrm{dict}\{S{:\,}[\textit{linear}, \textit{sag}, \textit{lbfgs}], P{:\,}[\textit{l2}]\}
\end{array}\]

The \emph{SMAC backend} implements the \pipe and \union combinators by
concatenating the hyperparameter dictionaries of the individual steps.
The SMAC backend adds a discriminant property $D$ into each dictionary
to track choices made for the \choice combinator, which it implements
via `$\vee$' branches.  It makes hyperparameter names unique by
adopting the scikit-learn name mangling convention of
\mbox{\textit{op}\python{\_\_}\textit{hp}}. Here is the generated SMAC
search space, eliding name mangling for readability:

\[\begin{array}{c@{\,}l@{\,}l@{\,}l@{\,}l@{}l}
     & \textrm{dict}\{N{:\,}(0..1),         & D{:\,}[\textit{J48}], & R{:\,}[\textit{false}],                                & C{:\,}(0..1)                     & \}\\
\vee & \textrm{dict}\{N{:\,}(0..1),         & D{:\,}[\textit{J48}], & R{:\,}[\textit{true},\textit{false}],                  & C{:\,}[0.25]                     & \}\\
\vee & \textrm{dict}\{N{:\,}[\textit{mle}], & D{:\,}[\textit{J48}], & R{:\,}[\textit{false}],                                & C{:\,}(0..1)                     & \}\\
\vee & \textrm{dict}\{N{:\,}[\textit{mle}], & D{:\,}[\textit{J48}], & R{:\,}[\textit{true},\textit{false}],                  & C{:\,}[0.25]                     & \}\\
\vee & \textrm{dict}\{N{:\,}(0..1),         & D{:\,}[\textit{LR}],  & S{:\,}[\textit{linear}],                               & P{:\,}[\textit{l1}, \textit{l2}] & \}\\
\vee & \textrm{dict}\{N{:\,}(0..1),         & D{:\,}[\textit{LR}],  & S{:\,}[\textit{linear}, \textit{sag}, \textit{lbfgs}], & P{:\,}[\textit{l2}]              & \}\\
\vee & \textrm{dict}\{N{:\,}[\textit{mle}], & D{:\,}[\textit{LR}],  & S{:\,}[\textit{linear}],                               & P{:\,}[\textit{l1}, \textit{l2}] & \}\\
\vee & \textrm{dict}\{N{:\,}[\textit{mle}], & D{:\,}[\textit{LR}],  & S{:\,}[\textit{linear}, \textit{sag}, \textit{lbfgs}], & P{:\,}[\textit{l2}]              & \}
\end{array}\]

The \emph{GridSearchCV backend} works similarly to the SMAC backend,
but adds one additional step. It discretizes each continuous
hyperparameter into a categorical. It accomplishes this by sampling a
user-configurable number of random values from the range and
distribution of the continuous hyperparameter. \lale adds a
distribution property in JSON Schema which is ignored by schema
validators but used by our compiler. Here is an excerpt of the
generated GridSearchCV search space:

\[\begin{array}{c@{\,}l@{\,}l@{\,}l@{\,}l@{}l}
     & \textrm{dict}\{N{:\,}[0.21,0.65,0.84], & D{:\,}[\textit{J48}], & R{:\,}[\textit{false}],                                & C{:\,}[0.07, 0.30, 0.89] & \} \vee \ldots\\
\vee & \textrm{dict}\{N{:\,}[\textit{mle}],   & D{:\,}[\textit{LR}],  & S{:\,}[\textit{linear}, \textit{sag}, \textit{lbfgs}], & P{:\,}[\textit{l2}] & \}
\end{array}\]

The \emph{hyperopt backend} takes advantage of the fact that hyperopt
supports fully nested search spaces. It adds a top-level `dict' over
steps of the \pipe and \union combinator, with a nested `$\vee$' over
choices of the \choice combinator. At the innermost level, it simply
includes per-step normalized schemas with discriminants~$D$. Here is
the generated hyperopt search space for our running example:

\[\textrm{dict}\left\{\begin{array}{c@{\,}l}
    0: & \textrm{dict}\{N{:\,}(0..1)\} \vee \textrm{dict}\{N{:\,}[\textit{mle}]\}\\
    1: & \left(\begin{array}{c@{\,}c}
                & \left(\begin{array}{c@{\,}l@{\,}l@{\,}l@{}l}
                         & \textrm{dict}\{D{:\,}[\textit{J48}], & R{:\,}[\textit{false}],               & C{:\,}(0..1) & \}\\
                    \vee & \textrm{dict}\{D{:\,}[\textit{J48}], & R{:\,}[\textit{true},\textit{false}], & C{:\,}[0.25] & \}
                  \end{array}\right)\\
           \vee & \left(\begin{array}{c@{\,}l@{\,}l@{\,}l@{}l}
                         & \textrm{dict}\{D{:\,}[\textit{LR}], & S{:\,}[\textit{linear}],                               & P{:\,}[\textit{l1}, \textit{l2}] & \}\\
                    \vee & \textrm{dict}\{D{:\,}[\textit{LR}], & S{:\,}[\textit{linear}, \textit{sag}, \textit{lbfgs}], & P{:\,}[\textit{l2}]              & \}
                  \end{array}\right)
         \end{array}\right)
\end{array}\right\}\]

Finally, in addition to a compiler that generates tool-specific search
spaces, \lale also provides a reverse mapping from a point in a
tool-specific search space back to a trainable \lale pipeline, which is 
essential for computing the loss being minimized by the CASH solver.
This 
reverse mapping uses the discriminant $D$ to select an algorithm from
the \choice operator. It then strips out the discriminant.
Furthermore, the reverse mapping interprets the
\mbox{\textit{op}\python{\_\_}\textit{hp}} name mangling to associate
hyperparameters with operators, then strips out the operator part and
configures the hyperparameter.


\section{Sklearn Compatible Portability}\label{sec:portability}

The auto-sklearn paper~\cite{feurer_et_al_2015} reported a few cases
where auto-Weka~\cite{thornton_et_al_2013} performed better. Often,
this was because ``the best classifier it chose is not implemented in
scikit-learn (trees with a pruning
component)''~\cite{feurer_et_al_2015}. Since scikit learn is
Python-based and Weka is Java-based, the two are non-trivial to use
side-by-side. \lale simplifies using
operators from different programming languages,
for different data modalities, and for DL (deep learning) and non-DL
models in the same pipeline.  Such mixed pipelines support both manual
(e.g.\ Figure~\ref{fig:lifecycle_notebook_manual}) and automated
(e.g.\ Figure~\ref{fig:lifecycle_notebook_auto}) machine learning.
Additionally, \lale provides familiar syntax for features in scikit
learn, making it easy for new users.  \lale pipelines can also be
passed to off-the-shelf scikit learn components such as cross validation or
GridSearchCV. Supporting them was challenging as they
expect specific behaviors of pipelines.

For a concrete example, we picked one of the cases where auto-Weka
outperformed auto-sklearn: the Car dataset~\cite{car_dataset} and the J48
operator, which is a tree with a pruning component. We also trained
various other operators on the same dataset: arulesCBA (classification
based on association rules~\cite{johnson_hahsler_2018}) from~R, a 
highly interpretable classifier;
XGBoost~\cite{chen_guestrin_2016}, a popular implementation of 
boosting with decision trees; and
a few scikit learn operators such as logistic regression.
This involved wrapping non-Python operators for use with \lale.  The
JSON Schema part worked the same irrespective of the programming
language: in all cases, it was straightforward to express side
constraints which our search space compiler could translate for CASH
tools.  For Java interoperability, we used the Python packages
\python{javabridge} and \python{weka}, then wrote a modest amount of
additional glue code to wrap \python{J48} for \lale. For R
interoperability, we used the Python package \python{rpy2}, then wrote
a wrapper
\python{ARulesCBAClassifier} for \lale. One tricky bit was
that arulesCBA uses R's lazy evaluation
feature~\cite{morandat_et_al_2012}, so we needed to create
uninterpreted \python{Formula} objects to pass across the
\python{rpy2} interface.

To enable using scikit learn components such as GridSearchCV on \lale
pipelines, \lale provides a compatibility wrapper for pipelines. This supports
the \python{sklearn.base.clone} and \python{set\_params} method. 
\python{sklearn.base.clone} is recursive and attempts to do all of its
work with \python{set\_params} if available, but \python{set\_params}
is impractical for cloning operators in Java or~R and for cloning some
advanced \lale features such as the \choice combinator.  The \lale
wrappers prevent this recursive behavior.


\section{Experiments}\label{sec:results}

This section demonstrates \lale's portability via different data
modalities and evaluates \lale's search space compiler.  The value
proposition of \lale is to leverage existing CASH tools effectively
and in a portable manner; in general, we do not expect \lale to
outperform them.

\begin{table}[!b]
  \caption{\label{tab:modalities_results}Performance of the best
    pipeline found using \lale with hyperopt. In all cases, the
    hyperopt trials used average 10-fold cross validation on the
    training set. For datasets that have a pre-defined train-test
    split, this table reports the accuracy on the test set, averaged
    over 3 experiments.}
  \centerline{\small\begin{tabular}[t]{llrr@{ }ll}
    Modality & Dataset & Iterations & \multicolumn{2}{c}{Accuracy (stdev)} & Accuracy type\\
    \midrule
    Text & Movie Reviews \cite{pang_lee_2005} & 100 & 77.20\% & (0.1) & 10-fold crossval\\
    Table & Car \cite{car_dataset} & 1000 & 98.07\% & (0) & test accuracy\\
    Image & CIFAR-10 \cite{Alex_cifar10_2009} & 50 & 93.53\% & (0.1058) & test accuracy \\
    Time-series & Epilepsy \cite{TUH_seizure_paper} & 50 & 73.15\% & (8.2) & test accuracy\\
  \end{tabular}}
\end{table}

To demonstrate portability, we picked four datasets from different
modalities. For each dataset, we specified a planned pipeline with
operator choices in \lale, and then used hyperopt to pick the best
operators and tune their hyperparameters.
Table~\ref{tab:modalities_results} summarizes the results.

\textbf{Text.}
For this modality we used the Movie Reviews dataset for binary
sentiment classification~\cite{pang_lee_2005}. The planned pipeline is
\python{(BERT | TFIDF) >{}> (LR | MLP | KNN | SVC | PAC)}, as seen in
Figure~\ref{fig:lifecycle_notebook_auto}. The best pipeline discovered
by hyperopt was \python{BERT >{}> LR}. While the accuracy of 77.2\%
does not match the state of the art, it is decent considering that
the classifier is a simple logistic regression.

\textbf{Table.}
For this modality we used the Car dataset~\cite{car_dataset},
consisting of structured data with categorical features, which we
label-encoded. The planned pipeline is \python{J48 | ARules | LR |
  KNN}, where \python{J48} from Weka~\cite{hall_et_al_2009} is
implemented in Java and \python{ARules} is implemented in
R~\cite{johnson_hahsler_2018}.  The best choice discovered by hyperopt
was \python{J48} from Weka, which means that portability paid off.

\textbf{Image.}
For this modality we used the CIFAR-10 computer vision
dataset~\cite{Alex_cifar10_2009}. We picked the
\python{ResNet50}~\cite{he_et_al_2015} deep-learning model,
since it has been shown to do well on CIFAR-10.  In our experiments,
we kept the architecture of \python{ResNet50} fixed, but varied
hyperparameters for learning procedure (number of epochs, batch size,
learning rate, and the type of learning rate decay).


\textbf{Time-series.}
For this modality we used the Epilepsy dataset, which is a subset of
the TUH Seizure Corpus \cite{TUH_seizure_paper}, for classifying
seizures by onset location (generalized or focal). 
We implemented a popular pre-processing method~\cite{Schindler2006} in
a \python{WindowTransformer} operator with three hyperparameters $W$,
$O$, and $T$. 
Note that this transformer leads to multiple samples per seizure.
Hence, during evaluation, each seizure is classified by taking a vote
of the predictions made by each sample generated from it.  The planned
pipeline is \mbox{\python{WindowTransformer >{}> (KNN | XGBoost | LR)
    >{}> Voting}}.  The transformer increases the number of rows, the
classifier works per-row, and the voting decreases the number of
rows. \lale pipelines can handle such non-trivial transformations 
and evaluations. 
The best pipeline discovered by hyperopt was
\python{WindowTransformer >{}> KNN >{}> Voting}.

\begin{wrapfigure}{r}{0pt}
\begin{minipage}{.8\textwidth}
\vspace*{-5mm}
\includegraphics[width=.5\textwidth]{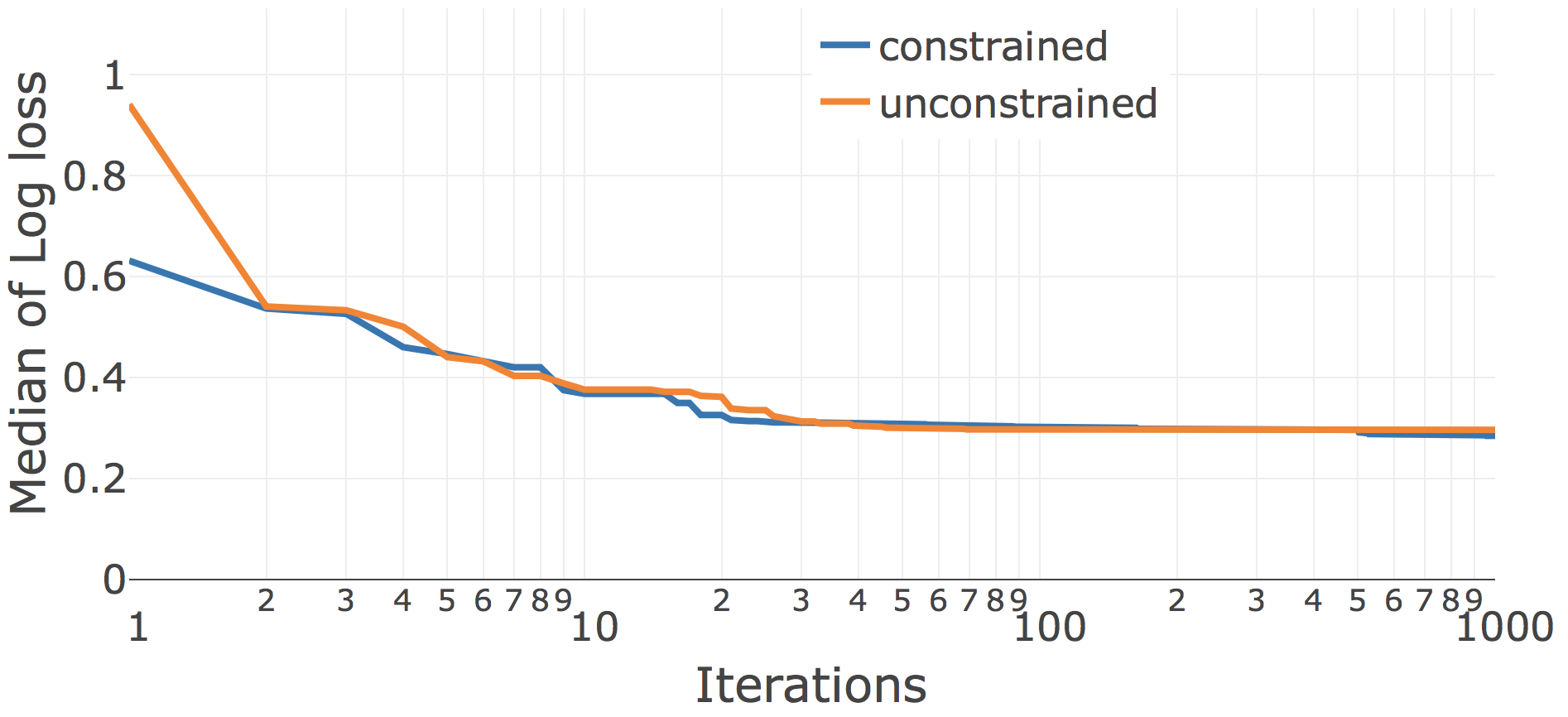}
\includegraphics[width=.5\textwidth]{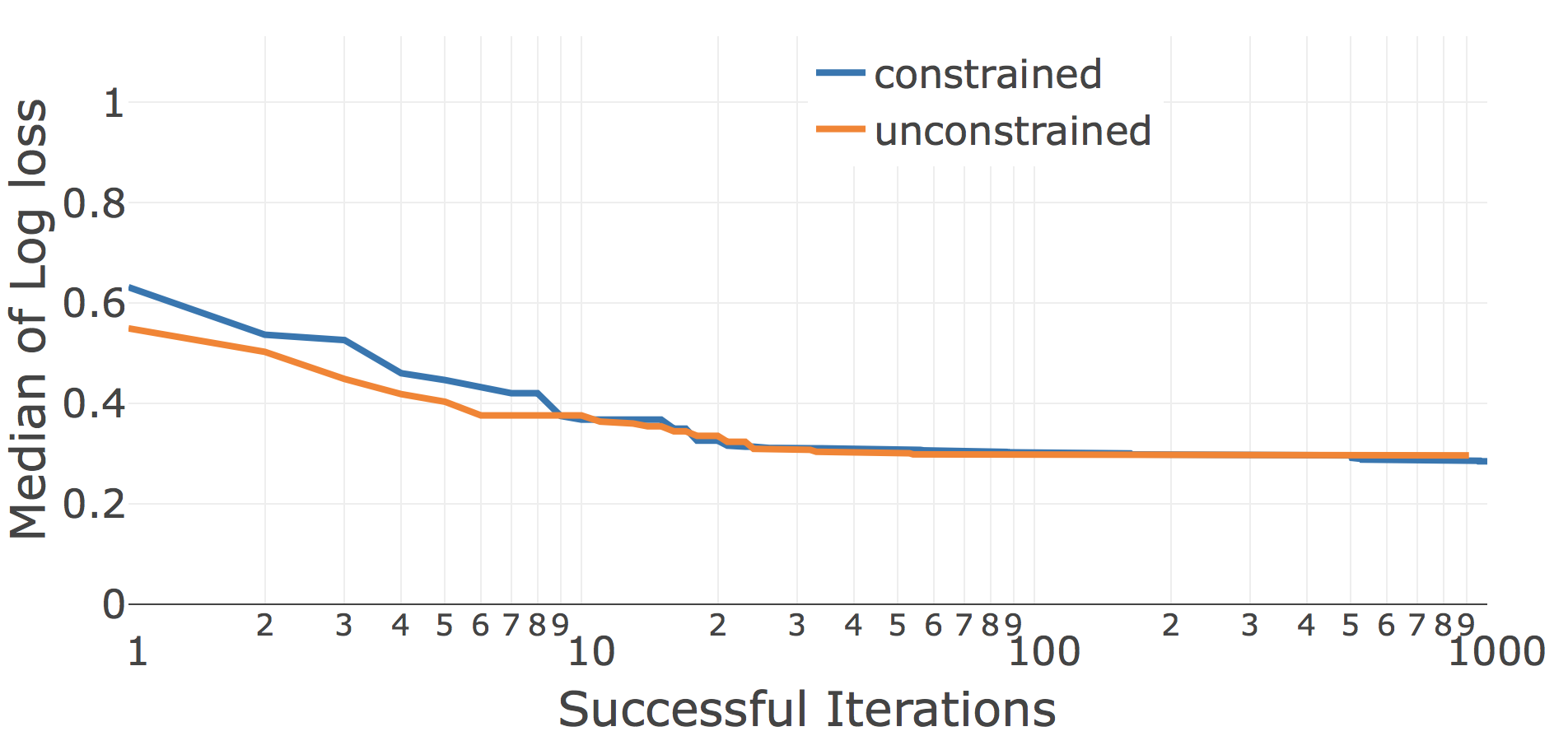}
\vspace*{-3mm}
\caption{\label{fig:car_2classifiers}Convergence with planned pipeline \python{LR | KNN}.}
\includegraphics[width=.5\textwidth]{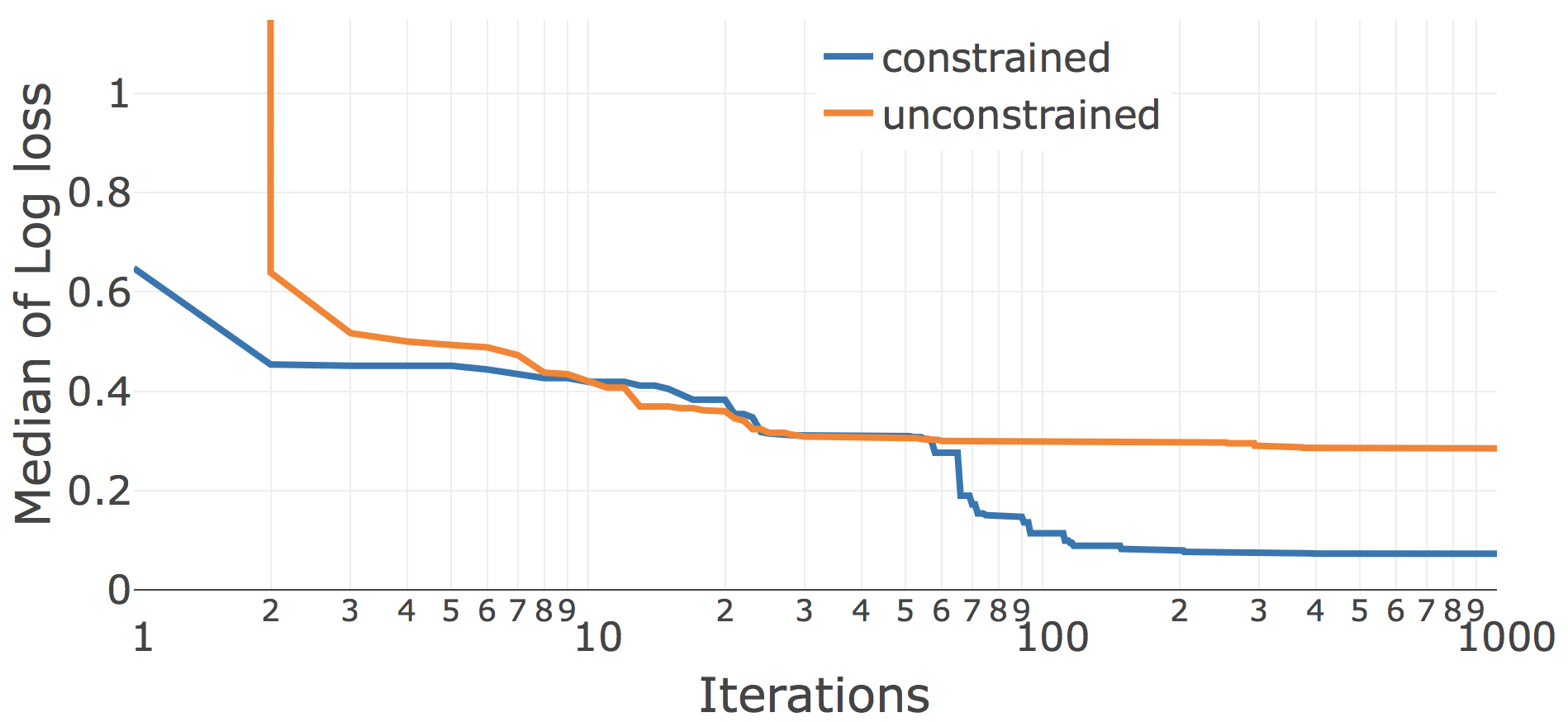}
\includegraphics[width=.5\textwidth]{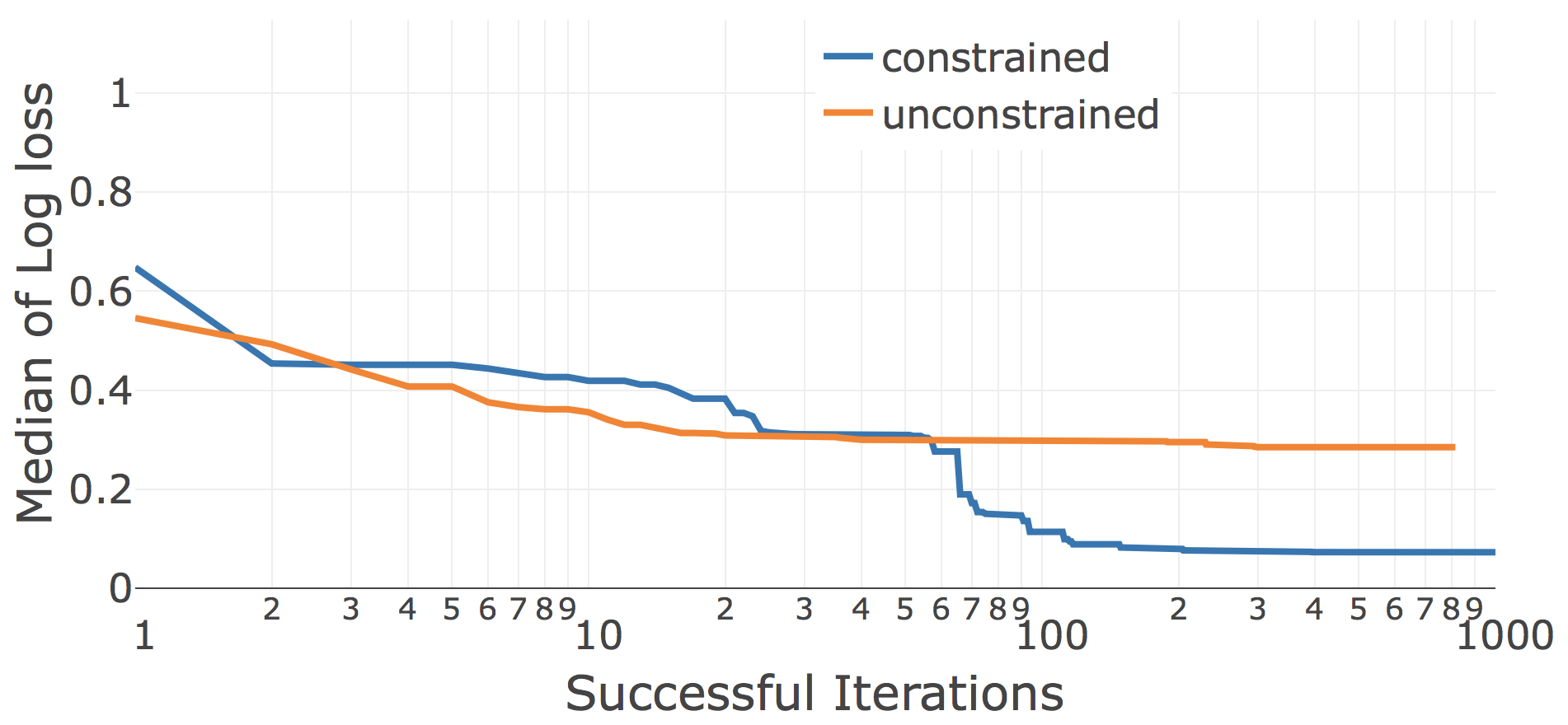}
\vspace*{-3mm}
\caption{\label{fig:car_3classifiers}Convergence with planned pipeline \python{J48 | LR | KNN}.}
\vspace*{-4mm}
\end{minipage}
\end{wrapfigure}
\textbf{Effect of side constraints on convergence.}
\lale's search space compiler takes rich hyperparameter schemas
including side constraints and translates them into semantically
equivalent search spaces for different CASH tools. This raises the
question of how important those side constraints are in practice.  To
explore this, we did an ablation study where we generated not just the
\emph{constrained} search spaces that are default with \lale but also
\emph{unconstrained} search spaces that drop side constraints. With
hyperopt on the unconstrained search space, some iterations are
unsuccessful due to exceptions, for which we reported
\python{np.float.max} loss.
Figure~\ref{fig:car_2classifiers} plots the convergence for the Car
dataset on the planned pipeline \python{LR | KNN}. Both of these
operators have a few side constraints. Whereas the unconstrained
search space causes some invalid points early in the search, the two
curves more-or-less coincide after about two dozen iterations.
The story looks very different in Figure~\ref{fig:car_3classifiers}
when adding a third operator \python{J48 | LR | KNN}.  In the
unconstrained case, \python{J48} has many more invalid runs, causing
hyperopt to see so many \python{np.float.max} loss values from
\python{J48} that it gives up on it. In the constrained case, on the
other hand, \python{J48} has no invalid runs, and hyperopt eventually
realizes that it can configure \python{J48} to obtain substantially
better performance.

\textbf{Results for different CASH tools.}
To explore whether the search spaces for different tools are indeed
equivalent, we generated search spaces from the planned pipeline
\python{J48 | ARules | LR | KNN} for both hyperopt and GridSearchCV,
then ran both tools on the Car dataset~\cite{car_dataset}.  We ran
hyperopt for 1000 iterations and GridSearchCV for 960 iterations
(there is no way to control the exact number of grid points in our
grid search implementation). Both CASH tools selected the \python{J48}
operator and converged to very similar but not completely identical
hyperparameter configurations for it.  Both CASH tools yielded the
same test accuracy of 98.07\%.


\section{Conclusion}\label{sec:conclusion}

This paper introduces \lale, a new Python DSEL for automated machine
learning. \lale resembles scikit
learn~\cite{buitinck_et_al_2013} while extending it for using CASH
tools such as SMAC~\cite{hutter_hoos_leytonbrown_2011} and
hyperopt~\cite{bergstra_et_al_2015}. It accomplishes this via three
novel contributions rooted in type systems: lifecycle as bindings,
types as search spaces, and scikit learn-compatible portability.
Collectively, these contributions help machine-learning practitioners
use more automation in a controlled and portable manner.

\bibliographystyle{plain}
\bibliography{bibfile}

\end{document}